\documentclass[12pt]{article}
\input epsf.tex

\begin{document}
\thispagestyle{empty}
\begin{flushright}
IMSc/2004/10/36 \\
DESY 04-203 \\
October 2004
\end{flushright}
\vskip 0.6in
\begin{center}
{\Large \bf An SO(10) GUT With See-Saw Masses For All Fermions \\}
\vskip .5in

\bf{\bf Bipin R. Desai${}^1$, G. Rajasekaran${}^2$ and Utpal Sarkar${}^{3,4}$}
\vskip .25in
{\sl ${}^1$ Physics Department, University of California, }\\
{\sl Riverside, CA 92521, USA}\\
{\sl ${}^2$ Institute of Mathematical Sciences, Chennai 600 113, India}\\
{\sl ${}^3$ Deutsches Elektronen-Synchrotron DESY, Hamburg, Germany}\\
{\sl ${}^4$ Physical Research Laboratory, Ahmedabad 380 009, India}

\end{center}
\vskip 1in

\begin{abstract}

\noindent

We propose an SO(10) grand unified theory which has the
simplest Higgs structure discussed so far in the literature.
We include only two Higgs scalars, a 210-plet and a 16-plet. 
In addition to the regular fermions we include one
singlet, whose mass term breaks chiral symmetry, so that
fermions can get masses. All fermions acquire see-saw masses,
since there are no Higgs bi-doublets. Required neutrino masses
with large mixing as well as leptogenesis are possible in this model. 

\end{abstract}
\newpage

The success of the standard model has motivated us to develop 
further our understanding of the basic interactions.
Although the only knowledge we have for physics beyond the
standard model comes from the neutrino mass, 
we have several theoretical motivations to
think about the extensions ahead of any experimental inputs. 
The most natural extension of the standard model is grand
unification, in which all the gauge coupling constants are
unified at some high energy and we have a simple unified 
gauge group governing all gauge interactions of nature. 
The simplest grand unified theory is based on the gauge 
group $SU(5)$. But the minimal $SU(5)$ GUT has some problems 
that involve fermion masses, proton decay, gauge coupling unification, 
etc. These problems may be eliminated in some extensions of
the model, but there is another compelling reason to consider
larger grand unified group. In $SU(5)$ the left-handed and
right-handed fermions are treated in different ways, so parity
is never conserved.

A natural extension of the standard model is the left-right
symmetric extension, in which parity is conserved at higher
energies and could be broken spontaneously \cite{lr}. The simplest
grand unified theory to accomodate the left-right symmetric
extension of the standard model is based on the gauge group
$SO(10)$. In recent times it has been noticed that in the
$SO(10)$ GUTs several interesting features come out 
naturally \cite{min1,min2,min3,min4,min5}. 
It has been realized that the minimal $SO(10)$ GUT has
considerable predictive power and thus attempts to construct GUTs
with simplest Higgs choice have become a challenging question
\cite{min1,min2,min3,min4}.

In all the left-right symmetric models one requires a 
bi-doublet Higgs scalar (which is a doublet under both the
left-handed and right-handed $SU(2)$ groups) 
to give masses to the fermions and break the electroweak 
symmetry. In a recent article it was pointed out that for
symmetry breaking the most economic choice is to consider
$SU(2)_R$ Higgs doublet to break left-right symmetry and
a $SU(2)_L$ doublet to break the electroweak symmetry
\cite{bms} and an explicit realization was then presented
in a supersymmetric model \cite{fre}. 
Even without a bi-doublet of Higgs scalar it
is possible to get fermion masses through see-saw 
contributions. Since a natural explanation of the smallness
of the neutrino masses requires see-saw mechanism
\cite{seesaw}, it will appear more natural if all fermion masses 
have same see-saw origin. See-saw contributions of fermion masses
were also studied extensively in the past with heavy
fermions \cite{ss1,ss2} to understand the flavor structure
of the fermions, but the present interest for see-saw
fermion mass is solely from the point of view of minimality
of the theory. In this article we intend to give
a realization of this idea of left-right symmetric models
without any bi-doublet Higgs scalars in an $SO(10)$ GUT with the
simplest possible Higgs structure considered so far. Most of 
the $SO(10)$ GUTs considers Higgs triplets \cite{typeII} to break the
left-right symmetry for several reasons \cite{min1,min2,min3,min5}.
However, Higgs doublets have also been considered for
the left-right symmetry breaking \cite{db}. In particular,
in superstring inspired models and in recent times in
orbifold GUTs Higgs doublets are the only choices \cite{orbi}. 

All grand unified theories have the gauge
hierarchy problem, which requires fine tuning of parameters
to maintain the electroweak symmetry breaking scale light.
This problem is solved by making the theory supersymmetric.
In supersymmetric theories although we need to make the
fine tuning at the tree level, there are no radiative
quadratic divergences which need to be fine tuned at 
each order of perturbation theory. Supersymmetric theories
are also advantageous because the cosmological constant
vanishes in the limit of exact supersymmetry. However,
in nature we have not seen any signals of supersymmetry
upto the electroweak symmetry breaking scale and 
at the same time we have
also found a non-vanishing cosmological constant. 
Supersymmetry cannot explain the smallness
of the cosmological constant and we may need to invoke
fine tuning to understand this.

Hence in recent times the question has arisen 
whether we really need supersymmetry \cite{split}. If we 
need fine tuning to understand the cosmological constant,
then perhaps we may need fine tuning to understand the gauge
hierarchy problem. It is hoped that some new physics at
very high energy will solve the fine tuning problems \cite{ft}
without invoking low energy supersymmetry. The advantages of
giving up supersymmetry at low energy are manifold since 
this gets rid of problems like the $\mu$ problem, CP
problem and flavor problem, which are all associated with low
energy supersymmetry. We shall thus not consider supersymmetry
in our construction at any stage and do not bother about
any fine tuning required to get any particular solution.

We first start with the different possibilities of constructing
an $SO(10)$ GUT without any 10-plet of Higgs scalars, which
contains a bi-doublet. Then we shall proceed to construct
the simplest possible model in terms Higgs representations.
The Higgs sector in our model is the smallest compared to all
the existing models of $SO(10)$ GUT. We introduce only two
Higgs scalars for all the symmetry breaking and for giving masses
to all the fermions including neutrinos. For chiral symmetry
breaking we have to introduce one heavy fermion singlet, but
otherwise there are no new ingredients.

In $SO(10)$ GUTs the fermions of each generation (including a 
right-handed neutrino) belong to the 16-plet spinor representation,
which transforms under the Pati-Salam subgroup ($G_{422} \equiv 
SU(4)_c \times SU(2)_L \times SU(2)_R$) as,
$$ 
\psi_{iL} \equiv {\bf 16 = (4,2,1) + (\bar 4, 1, 2)} .
$$
$i = 1,2,3$ is the generation index. The right-handed fermions 
($\psi_{iR}$) then belong to the conjugate representation, 
$$\psi_{iR} \equiv {\bf \bar{16} = (\bar 4,2,1) + ( 4,1,2)} .$$ 
The generators of the left-right symemtric group 
$G_{3221} \equiv SU(3)_c \times SU(2)_L \times SU(2)_R 
\times U(1)_{B-L}$ are then related to the electric charge by
$$ Q = T_{3L} + T_{3R} + {(B-L) \over 2} = T_{3L} + {Y \over 2} .
$$ 
The quarks and leptons then transform as
\begin{eqnarray}
(4,2,1) &=& \left\{ \begin{array}{l} 
q_L = \pmatrix{ u&d}_L \equiv (3,2,1,1/3) \cr
\ell_L = \pmatrix{ \nu & e }_L \equiv (1,2,1,-1) \end{array}
\right . \nonumber \\ && \nonumber \\
(\bar 4,1,2) &=& \left\{ \begin{array}{l} 
{q_R}^c = {q^c}_L = \pmatrix{d^c & u^c}_L \equiv (\bar 3,1,2,-1/3) \cr
{\ell_R}^c = {\ell^c}_L = 
\pmatrix{e^c & \nu^c }_L \equiv (1,1,2,1) \end{array} \right .
\end{eqnarray}
where (x,y,z) and (x,y,z,w) denote the transformation property
under $G_{422}$ and $G_{3221}$ respectively.
Similarly the right-handed fermions belong to $\overline{16}$. 

The minimal Higgs representation required to break the groups
$SU(2)_L$ and $SU(2)_R$ are two Higgs doublets, both of which
belong to a ${\bf 16}$-plet Higgs representation $\Gamma$.
There are two neutral components of $\Gamma$,
\begin{eqnarray}
\chi_L & \equiv & (1,2,1,-1) \subset (4,2,1) \subset 16
\equiv \Gamma \nonumber \\
{\chi^c}_L & \equiv & (1,1,2,1) \subset (\bar 4,1,2) \subset 16
\equiv \Gamma \nonumber 
\end{eqnarray}
and the corresponding conjugates belonging to $\Gamma^\dagger$,
\begin{eqnarray}
\chi_R & \equiv & (1,1,2,-1) \subset (4,1,2) \subset \overline{16}
\equiv \overline{\Gamma} \nonumber \\
{\chi^c}_R & \equiv & (1,2,1,1) \subset (\bar 4,2,1) \subset 
\overline{16} \equiv \overline{\Gamma}. \nonumber
\end{eqnarray}
The vacuum expectation values ($vev$s) of the 
Higgs fields $\chi_R$ and ${\chi^c}_L$ can break
the left-right symmetry to the standard model 
($G_{3221} \to G_{321} \equiv S(3)_c \times SU(2)_L \times U(1)_Y$)
and the $vev$s of $\chi_L$ and ${\chi^c}_R$ can break
the electroweak symmetry. However, these fields cannot give
masses to the fermions since chiral symmetry is not broken
by any of these Higgs scalars. 

There is one more problem with these Higgs scalars, which has
to be taken care of. Usually parity is conserved in most
left-right symmetric models, so that spontaneous breaking of 
parity can give an explanation of the origin of parity violation. 
It is also possible to start with a theory, in which parity is
explicitly broken so that the left-handed gauge couplings could
be different from the right-handed gauge couplings. 
In theories with conserved left-right parity, it is difficult to
give different and non-zero vevs to the fields $\chi_L$ and $\chi_R$.
The minimization of the potential leads to either equal values of these
vevs or zero value for at least one of them. To solve this
problem one may consider an extra $O(2)$ symmetry, which is
most unnatural \cite{fs}. A natural solution to this problem
is obtained by breaking parity. For any low energy theories,
it is convenient to start with an explicit parity violation
and hence with different left-handed and right-handed couplings.
However, since we are starting from an $SO(10)$ GUT which
includes parity as one of its generators, we shall consider a more
appealing scenario in which parity is spontaneously broken
before the left-right symmetry breaking by a parity odd
singlet field \cite{par}.

In $SO(10)$ GUTs, the discrete left-right parity is a 
generator of the group, which is referred to as $D-$parity \cite{par}. 
As a result, to start with, parity is always conserved. 
So starting from $SO(10)$ it is not possible
to construct any model with explicit parity violation. However,
it is possible to break parity at very high energy spontaneously, 
before the breaking of $SU(2)_R$, which will then allow us to have
$\langle \chi_L \rangle \neq \langle  \chi_R \rangle$. This is
done by giving a $vev$ to some $D-$parity odd field. Before
we proceed further, let us discuss the other Higgs scalar in
the model.

We also need one more Higgs scalar to break the $SO(10)$ group
to its left-right symmetric subgroup. Although there are more
than one possibilities, we choose the ${\bf 210}$ dimensional
representation, since it also serves another purpose which
we shall discuss later. The ${\bf 210}$ fields decompose under
the Pati-Salam subgroup ($G_{422}$) as,
\begin{eqnarray}
\Phi \equiv {\bf 210} &=& {\bf (1,1,1) + (6,2,2) + (15,3,1) + (15,1,3)}
\nonumber \\ && {\bf 
+ (15,1,1) + (10,2,2) + (\overline{10},2,2)}
\nonumber
\end{eqnarray}
The field $\Phi$ breaks the group $SO(10)$ to $G_{3221}$ when
the components $(1,1,1)$ and $(15,1,1)$ acquire 
$vev$s at the GUT scale $M_U$. The component $(1,1,1)$ is odd
under D-parity of the group $SO(10)$. As a result, this allows
the left-handed neutral components of $\Gamma$ to become light
while keeping the right-handed components heavy. This parity
odd singlet plays a crucial role in giving neutrino masses.
In the absence of this parity odd field the lightest neutrino
always remain massless. 

Since this is crucial for our discussion, we shall ellaborate 
this point. Consider the $SO(6) \times SO(4)$ subgroup of $SO(10)$,
which is also the Pati-Salam subgroup since $SO(6)$ is isomorphic
to $SU(4)$ and $SO(4)$ is isomorphic to $SU(2) \times SU(2)$. 
The ${\bf 210}$ representation is a totally antisymmetric
tensor of rank four $\Phi_{abcd}$ and the singlet $(1,1,1)$ is the 
component $\Phi_{6789}$ in the notation in which $a,b,c,d = 0,1,..,5$
are $SO(6)$ indices and $a,b,c,d = 6,7,8,9$ are $SO(4)$ indices. 
It can be shown that the action of the $D-$parity operator on
this field gives $-1$ and hence it is odd. However, for an easy
understanding we shall give a simple argument showing why this 
is odd under $D-$parity. There are two singlets in 
${\bf 16 \times \overline{16} = 1 + 45 + 210}$ 
which are, $A= (4,2,1) \times (\bar 4,2,1)$ and $B =
(\bar 4,1,2) \times (4,1,2)$. Under $D-$parity $A \leftrightarrow B$,
and hence any of these singlets $A$ or $B$ cannot be the singlet
of $SO(10)$, since $D-$parity is a generator of $SO(10)$. So,
the combination $A+B = 1$ of $SO(10)$ and the
$A-B = (1,1,1) \subset 210$ and hence under $D-$parity, 
$\Phi_{6789} \leftrightarrow - \Phi_{6789}$. 

We shall now come back to the Higgs doublets and show how 
the D-parity breaking will make the fields $\chi_L$ and $\chi^c_R$
as light as 100 GeV, so that $vev$s of these fields can break 
the electroweak symmetry $\langle \chi_L \rangle = u_L \sim 100$ GeV. 
The right-handed components $\chi_R$ and $\chi_L^c$ will
remain almost as heavy as the GUT scale $\langle \chi_R \rangle 
= u_R \sim M_R \sim 10^{14}$ GeV. This requires fine 
tuning, but as we argued at the begining we allow it. 
The D-parity breaking also ensures the inequality of the
$vev$s $u_L$ and $u_R$. 

We shall first write down the potential for the scalar fields
in this model,
\begin{eqnarray}
{\cal L}_s &=& m_\Phi^2 \Phi^2 + \eta \Phi^3 + {\lambda_\Phi \over
4!} \Phi^4 + m_\Gamma^2 \Gamma^\dagger \Gamma + {\lambda_\Gamma
\over 4} (\Gamma^\dagger \Gamma)^2 \nonumber \\
&+& {\lambda_\Gamma^\prime \over 4} [\Gamma^4 + (\Gamma^\dagger)^4]
+ M_D \Phi ( \Gamma^\dagger \Gamma)  + \lambda_{\Phi \Gamma}
\Phi^2 ( \Gamma^\dagger \Gamma)
\end{eqnarray}
The coupling $\Phi \Gamma \Gamma^\dagger$ is the crucial
term, which breaks the D-parity when the singlet component (1,1,1)
of Phi acquires nonvanishing vev and gives the mass splitting
between $\chi_L$ and $\chi_R$. The $vev$s $u_L$ and $u_R$ also
split and the lightest neutrino gets mass due to this term. 

We shall now discuss the masses of the components of $\Gamma$
and the $vev$s. The singlet component of $\Phi$ (1,1,1) is
odd under the D-parity, which is the parity operator acting on
the $SO(10)$ group space. If we denote this component as
$\eta = \Phi (1,1,1)$, then the scalar potential responsible
for the masses of the fields $\chi_L$ and $\chi_R$ is given by,
\begin{eqnarray}
V &=& m_\Gamma^2 ( \chi_L^c \chi_R + \chi^c_R \chi_L)
+ M_D ~ \eta (\chi_L^c \chi_R - \chi^c_R \chi_L)
\nonumber \\
&+& \lambda_{\Phi \Gamma} ~\eta^2 (\chi_L^c \chi_R + \chi^c_R \chi_L).
\end{eqnarray}
The masses of these fields are then given by,
\begin{eqnarray}
\mu_L^2 &=& m_\Gamma^2 - M_D \langle \eta \rangle + \lambda_{\Phi \Gamma} 
 \langle \eta \rangle^2 ,\nonumber \\
\mu_R^2 &=& m_\Gamma^2 + M_D \langle \eta \rangle + \lambda_{\Phi \Gamma}
 \langle \eta \rangle^2 .
\end{eqnarray}
With proper fine tuning we now get,
\begin{eqnarray}
\langle \chi_L \rangle &=& \langle \chi^c_R \rangle = u_L 
\sim \mu_L \sim 100~{\rm GeV}
\nonumber \\
\langle \chi_R \rangle &=& \langle \chi^c_L \rangle = u_R 
\sim \mu_R \sim M_U \gg u_L .
\end{eqnarray}
Thus the two components of the field $\Gamma$ acquire widely
different masses and $vev$s. $u_L$ breaks the electroweak
symmetry, while $u_R$ breaks the left-right symmetry at a very
high scale, close to the GUT scale. 

This demonstrates how the fields $\Phi$ and $\Gamma$ are
sufficient to break the group $SO(10)$ to the left-right subgroup
at the GUT scale and then at some intermediate scale break the
left-right symmetric model to the standard model and subsequently
break the electroweak symmetry. We shall now discuss how one can
give masses to the fermions without introducing any new Higgs scalars,
although we may introduce new fermions, which may acquire
masses at very high scale breaking the chiral symmetry. 

We shall first consider only tree level mass generation. 
Let us start with a list of  
dimension-5 operators, which can give Dirac masses to the
quarks and leptons and Majorana masses to the left-handed
and right-handed neutrinos. They are,
$$
\begin{array}{rclcrcl}
{\cal O}_1 &=& (\overline{q_L} {\chi_L})(q_R {\chi_R}^c) & ~~~~&
{\cal O}_2 &=&  (\overline{q_L} {\chi_L}^c)(q_R {\chi}_R) \cr
{\cal O}_3 &=&  (\overline{\ell_L} {\chi_L})(\ell_R {\chi_R}^c) &&
{\cal O}_4 &=&  (\overline{\ell_L} {\chi_L}^c)(\ell_R {\chi}_R) \cr
{\cal O}_5 &=&  (\ell_L {\chi_L}^c)(\ell_L {\chi_L}^c) &&
{\cal O}_6 &=&  (\ell_R {\chi_R}^c)(\ell_R {\chi_R}^c)
\end{array}
$$
${\cal O}_1$ and ${\cal O}_2$ contributes to up and down quark masses, 
${\cal O}_3$ gives Dirac masses to the neutrinos, 
${\cal O}_4$ contributes to charged lepton masses
and ${\cal O}_5$ and ${\cal O}_6$ are the Majorana masses for the
left-handed and right-handed neutrinos. 

These operators may be realized in three ways. A scalar field
may mediate these terms, in which case we are back to the conventional
models since the intermediate scalars have to be either ${\bf 10,
120}$ or ${\bf 126}$. The next possibilities are when these 
effective terms are generated by intermediate fermions. In this case
we need two fermions for each of the first four operators, one left-handed
fermion and one right-handed fermions with same quantum numbers, while
for the last two operators we may not require any new fermions. 
With only one fermion it is not possible to generate any of the
terms, since that will not break the chiral symmetry. So, we have
to introduce two fields with same quantum numbers but opposite
chirality and allow the mass terms of these fields in the Lagrangian
which will break the chiral symmetry. 

For ${\cal O}_1$ we need one left-handed and one right-handed fields
both transforming as $U_{L,R} \equiv (3,1,1,4/3) \subset (15,1,1) 
\subset 45 ~~{\rm or}~~ 210$. Then the essential terms in the 
Lagrangian which can allow ${\cal O}_1$ are given by,
\begin{equation}
{\cal L}_1 = a_1 \overline{U_L} q_R {\chi_R}^c + b_1
\overline{q_L} U_R {\chi_L} + m_U \overline{U_L} U_R + H.c.
\end{equation}
Similarly, for the operator ${\cal O}_2$ we need the fields
$D_{L,R} \equiv (3,1,1,-2/3) \subset (6,1,1) \subset 10 ~~{\rm or}~~ 
126$ or $D_{L,R} \equiv (3,1,1,-2/3) \subset (10,1,1) \subset 120$.
For the operator ${\cal O}_4$ we need $E_{L,R} \equiv 
(1,1,1,-2) \subset (10,1,1) \subset 120$. The neutrino sector is 
somewhat simpler and we can manage with  
$S_{L,R} \equiv (1,1,1,0) \subset (1,1,1) \subset 1~~{\rm or}~~ 45$. 
Thus the simplest possibility will be to have one $120$ to give
masses to the down quarks and charged leptons and one $45$ to give
masses to the up quarks and the neutrinos. 

\begin{figure}[h!]
\begin{center}
\epsfxsize8cm\epsffile{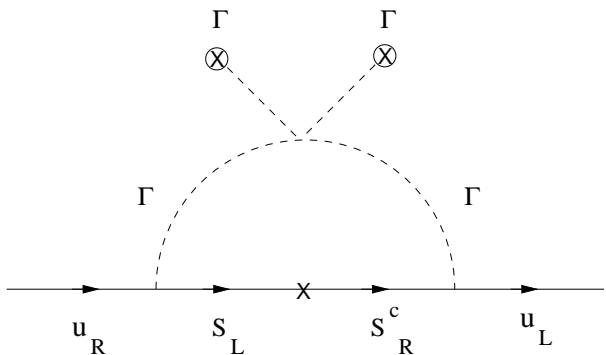}
\caption{One loop diagram contributing to the fermion masses.}
\label{xyfig1}
\end{center}
\end{figure}

We shall now discuss the third and most interesting possibility
with radiative generation of the fermion masses. We now include 
one heavy $SO(10)$ singlet fermions $S_{aL} \equiv {\bf 1 = (1,1,1)}$
per generation ($a=1,2,3$). The most general Yukawa couplings are 
then given by,
\begin{eqnarray}
{\cal L}_Y &=& f \overline{S^c_R} \psi_L \Gamma^\dagger
+  f \overline{S_L} \psi_R \Gamma + M_S S_L S_L + M_S S^c_R S^c_R .
\end{eqnarray}
We have written the Hermitian conjugate term separately for
clarity. Since we are not discussing the question of CP violation,
we assume all couplings are real. 
Chiral symmetry is broken by the mass term of the
singlet $M_S$. We also assume that the
scale of chiral symmetry breaking, $M_S$ is close to the GUT scale.

\begin{figure}[ht!]
\begin{center}
\epsfxsize8cm\epsffile{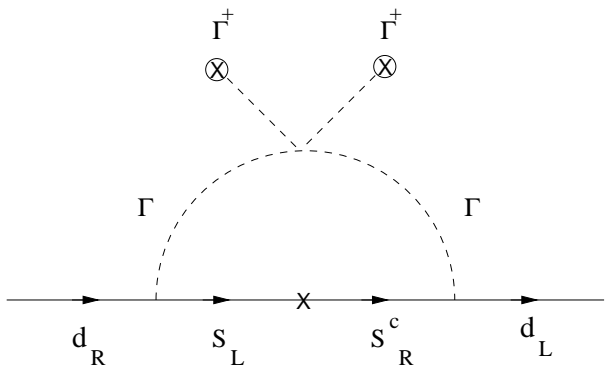}
\caption{One loop diagram contributing to the fermion masses.}
\label{xyfig2}
\end{center}
\end{figure}

Once chiral symmetry is broken,
fermions can get masses through one loop diagrams of figure 1 and
figure 2. The diagram in figure 1 generates the effective operators
${\cal O}_1$ and ${\cal O}_3$, which are of the form
\begin{equation} 
\overline{\psi_L}~ \psi_R ~ \chi_L ~{\chi^c}_L \subset
\overline{\psi_L}~ \psi_R ~ \Gamma ~\Gamma. 
\label{xx} \end{equation}
On the other hand the diagram in figure 2 
generates the effective operators ${\cal O}_2$ and ${\cal O}_4$,
which are of the form
\begin{equation} 
\overline{\psi_L}~ \psi_R ~ \chi_R ~{\chi^c}_R \subset
\overline{\psi_L}~ \psi_R ~ \Gamma^\dagger ~\Gamma^\dagger. 
\end{equation}
Since the term $(\Gamma^\dagger \Gamma)^2$ enters in figure 1,
while $(\Gamma^4 + {\Gamma^\dagger}^4)$ enters in figure 2,
the up and down quark masses are not the same. The
question of neutrino masses is somewhat complicated and we
shall discuss it later.

We can now write down the fermion masses in this model. 
For convenience we define, $$ m_L = f u_L , ~~~~~
{\rm and} ~~~~~ m_R = f u_R .$$
The diagram of figure 1 generates the up quark mass,
\begin{equation}
M_u = {\lambda_\Gamma \over 8 \pi} { m_R m_L \over M_X}
\end{equation}
and figure 2 generates the down quark and charged lepton
masses,
\begin{equation}
M_{d,\ell} = {\lambda_\Gamma^\prime \over 8 \pi} { m_R m_L \over M_X} .
\end{equation}
Here $M_X = M_\Gamma^2 /M_S$ or $M_S$, depending on whether
$M_\Gamma$ or $M_S$ is larger. 

The up and down quark
mass differences are explained by the different coupling
constants $\lambda_\Gamma$ and $\lambda_{\Gamma}^\prime$. 
Although this also gives the $b-\tau$ unification, it does
not give us the right fermion mass relations for the first
and second generations. We hope that some new physics near
the GUT scale can solve this problem. For example, if this 
$SO(10)$ GUT descends from a $E_6$ GUT, then the fundamental
representation of $E_6$ will contain a $10-$plet and a
singlet of fermion. The $10-$plet fermion can now be very
heavy and can contribute to only the down quark sector, 
solving this fermion mass problem. In fact, if there are 
heavy fermions $D_{L,R}$ or $E_{L,R}$ in the representations 
120 or 126, they can
also solve this fermion mass problem. There could be other
particles in the loop, which can contribute differently to
the fermions solving this problem which was discussed in
some of the earlier references \cite{ss1,ss2}. 

We shall now come to the question of neutrino masses.
Neutrino masses with doublets and singlets have already
been studied in the literature \cite{min4,db,neut}. However, 
in the earlier papers Higgs bi-doublet was present and
our scenario without any Higgs bi-doublet 
has a special feature due to ${\cal D}$ parity
which we shall explain. 
Although there are radiative corrections to the neutrino
masses, we may neglect them in comparison to the tree level
contributions. For completeness of our discussions
we shall consider them. The neutrinos will now mix with the 
singlet fermion $S_L$.
Although there are no mass terms for the neutrinos, due to 
this mixing neutrinos will get an induced mass. We can now write
down the mass matrix in the basis $\pmatrix{ \nu_L & {\nu^c}_L &
S_L }$,
\begin{equation}
M_\nu = \pmatrix{ 0 & 0 & m_L \cr 0 & 0 &m_R  \cr
m_L & m_R & M_S } .
\end{equation}
Diagonalization of this matrix will give two heavy states $S_L$
and ${\nu^c}_L$ with eigenvalues $M_S$ and $m_R^2/M_S$, but the
left-handed neutrinos will remain massless. 
One loop contributions do not solve this problem since they are
also proportional to the effective contributions one can
get after integrating out the heavy singlet field. As a
result the left-handed neutrinos still remain massless. One may
try to extend the theory with two or more singlets, but even then 
the determinant of the mass matrix vanishes and the lightest 
left-handed neutrino cannot get any mass. Apparently the 
left-right ${\cal D}$ parity symmetry makes the determinant of
the neutrino mass matrix to vanish as we shall discuss next. 

The effective neutrino Dirac mass term comes from the operator of
equation \ref{xx} and including the Majorana masses the
effective operator can be written as
\begin{equation} 
\nu_L~ \nu_L~ {\chi^c}_R~ {\chi^c}_R + {\nu^c}_L~ {\nu^c}_L~ 
\chi_R ~\chi_R + {\nu^c}_L~ \nu_L ~\chi_R ~{\chi^c}_R \subset
\psi_L~ \psi_L ~ \Gamma^\dagger ~\Gamma^\dagger  + H.c.. 
\label{xx1} \end{equation}
Thus the effective neutrino mass term in
the basis $\pmatrix{\nu_L & {\nu^c}_L}$ takes the form
(with $\beta$ taken to be some effective coupling constant)
$$ M_\nu = {\beta \over M_S} \pmatrix{ m_L^2 & m_L m_R 
\cr m_L m_R & m_R^2} $$
whose determinant vanishes because the coupling constants are the
same for $\nu_L$ and $\nu^c_L$, implying a massless left-handed
neutrino, since $u_R \gg u_L$. It should be noted that this 
is also related to the $D-$parity. Thus if we can include
the $D-$parity odd singlet $\Phi$ in this effective neutrino 
mass operator, then this problem may be solved. We do this by
combining the effective operator of equation \ref{xx1} with
the effective operator
\begin{equation} 
\psi_L~ \psi_L ~ \Gamma^\dagger ~\Gamma^\dagger ~\Phi + H.c \supset
- \nu_L~ \nu_L~ {\chi^c}_R~ {\chi^c}_R ~\eta + {\nu^c}_L~ {\nu^c}_L~ 
\chi_R ~\chi_R ~\eta .
\label{xx2} \end{equation}
Since the field $\eta$ is odd under parity and the first
term goes to the Hermitian conjugate of the second term, 
there is a $(-ve)$ sign in the first 
term. In addition, the Dirac mass now disappears since it
is even under D-parity while $\eta$ is odd. 

The effective operators of equation \ref{xx1} and equation
\ref{xx2} could come from the figures 3(a) and 3(b) respectively.
\begin{figure}[ht!]
\begin{center}
\epsfxsize12cm\epsffile{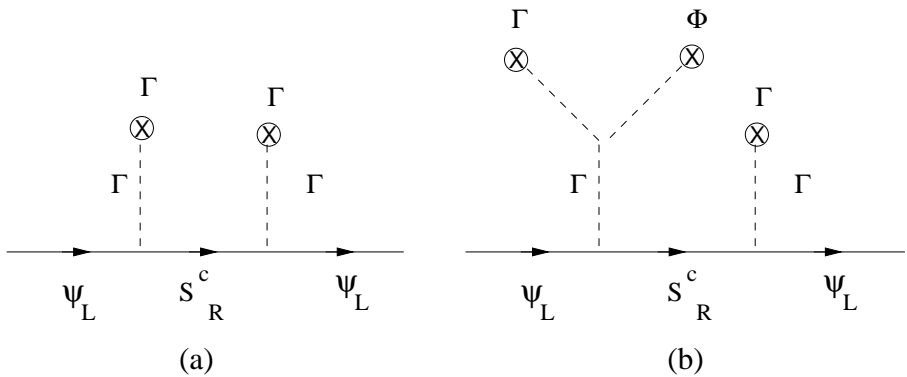}
\caption{Tree level diagrams contributing to the neutrino masses.}
\label{xyfig3}
\end{center}
\end{figure}
Including both these contributions we can now write down the
neutrino mass matrix as,
\begin{equation} 
M_\nu = \pmatrix{ (1-\alpha) {m_L^2 \over M_S} & {m_L m_R 
\over M_S} \cr {m_L m_R \over M_S} & (1+\alpha) {m_R^2
\over M_S}} .
\end{equation}
where, $ \alpha = M_D ~{\langle \Phi \rangle / m_\Phi^2}$.
In the limit $\alpha >-1$ and $m_L \ll m_R$, 
diagonalization of this matrix gives a heavy neutrino ${\nu^c}_L$
with mass of the order of $m_R^2/M_S$ and the left-handed
neutrino remains light with mass
\begin{equation} 
m_\nu =  { \alpha^2 \over (1 + \alpha)}  {m_L^2 \over M_S} .
\end{equation}
With $m_L$ to be of the order of electroweak symmetry breaking 
scale and $M_{S}$ to be of the order of the GUT scale, 
this gives the required neutrino mass to be fraction of an eV. 
It is to be noted that all fermion masses are of see-saw type. 
Since the Dirac masses are of the form, $m_L m_R/M_S$, they can
be as heavy as top quark masses, while the Majorana mass for the
left-handed neutrinos is of the form $m_L^2/M_S$ 
and hence remain very light and of the order of fraction of eV.

Since the neutrino masses now depend on the couplings with
the singlets, there is no stringent restriction coming from the up
quark masses. As a result, it may be possible to get large 
neutrino mixing angles. The right-handed neutrinos and the new singlet
fermions can now decay into light leptons. The Majorana masses
of the left-handed and right-handed singlets violate lepton
numbers, which in turn can generate enough lepton asymmetry.
Before the electroweak phase transition this asymmetry can then
generate a baryon asymmetry of the universe \cite{fy}. Since there is no
supersymmetry, the gravitino bounds are not present. 
The out-of-equilibrium condition can be satisfied near the GUT
scale since the couplings are large to get the required neutrino
mass with large see-saw scale. In this model there is another
interesting feature that the singlets combine with the right-handed
neutrinos to form pseudo-Dirac particles and hence resonant
leptogenesis is also possible \cite{rlepto1,rlepto2}. We shall present all 
these details in a forthcoming article \cite{next}.

In conclusion, we constructed an $SO(10)$ GUT without any Higgs
bi-doublets. All the symmetry breaking could be achieved by only
two Higgs scalars, a ${\bf 210}$ and a ${\bf 16}$. By including a
massive singlet fermion we break chiral
symmetry which can then give masses to all the fermions radiatively
without introducing any new scalar fields. All fermion
masses have the same see-saw form. The spontaneous parity
breaking plays a crucial role in breaking the left-handed
and right-handed $SU(2)$ groups
at two widely different scales and also giving masses
to the left-handed neutrinos in this scenario. Large neutrino
mixing and leptogenesis is also possible in this scenario. 

{\sl Acknowledgement}
We thank Prof E. Ma and Dr M. Frigerio for discussions.
One of us (US) would like to thank Prof W. Buchmuller for inviting
him to visit DESY and acknowledge hospitality at DESY.
GR acknowledges the support of the DAE-BRNS Senior Scientist Scheme and
the hospitality of the Physics Department,University of California,
Riverside. BRD acknowledges
the support in part by the U.S. Department of Energy under 
Grant No. DE-FG03-94ER40837.

\newpage


\end{document}